
\documentstyle{amsppt}
\mag=\magstep1
\nopagenumbers
\NoRunningHeads

\def\sig{\sigma}
\def\lam{\lambda}
\def\tens{\otimes}
\def\Comp{{\Bbb C}}
\def\Int{{\Bbb Z}}
\def\tr{\operatorname{tr}}
\def\half{{\textstyle \frac12}}

\topmatter
\title
Generalized $XYZ$ Model \\
Associated to Sklyanin Algebra$^*$
\endtitle

\author
Takashi Takebe
\endauthor

\affil
Department of Mathematical Sciences \\
the University of Tokyo\\
Hongo 7-3-1, Bunkyo-ku, Tokyo 113, Japan
\endaffil

\abstract
The free energy of a lattice model, which is a generalization
of the Heisenberg $XYZ$ model with the higher spin representation
of the Sklyanin algebra, is calculated by the generalized
Bethe Ansatz of Takhtajan and Faddeev.
\endabstract

\endtopmatter

\document

\subhead \S0. Introduction \endsubhead

\footnote""{$*$ talk given at the XXI Differential
Geometry Methods in Theoretical Physics, Tianjin,
China 5-9 June 1992 ; hep-th/9210086}
The $XYZ$ model, a quantum chain system with Hamiltonian
$$
H = -\frac12 \sum_{n=1}^N (
J_x \sig_n^x \sig_{n+1}^x +
J_y \sig_n^y \sig_{n+1}^y +
J_z \sig_n^z \sig_{n+1}^z ),
$$
was first solved by Baxter in early 70's by means of the
Bethe Ansatz. This work was interpreted from the view point
of the quantum spectral transformation method (QSTM) by
Takhtajan and Faddeev in 1979 [TF]. Their starting point is
the fundamental relation of the $L$ operators
$$
R^{12}(\lam-\mu) L^{13}(\lam) L^{23}(\mu) =
                 L^{23}(\mu)  L^{13}(\lam) R^{12}(\lam-\mu),
\tag0.1
$$
where
$$
\gather
L(\lam) = \sum_{a=0}^3 W_a^L(\lam) \sig^a \tens \sig^a,
\tag0.2
\\
W_0^L(\lam) = \frac{\theta_{11}(\lam)}{\theta_{11}(\eta)},
W_1^L(\lam) = \frac{\theta_{10}(\lam)}{\theta_{10}(\eta)},
W_2^L(\lam) = \frac{\theta_{00}(\lam)}{\theta_{00}(\eta)},
W_3^L(\lam) = \frac{\theta_{01}(\lam)}{\theta_{01}(\eta)},
\endgather
$$

$\sig^a$'s are Pauli matrices,
$\theta_{ab}(\lam) = \theta_{ab}(\lam;\tau)$ (we follow
the notations of [M] for the theta functions), and
the $R$ matrix is the Baxter's $R$ matrix defined by
$$
R(\lam) = \sum_{a=0}^3 W_a^R(\lam) \sig^a \tens \sig^a,
W_a^R(\lam) := W_a^L(\lam + \eta).
$$
Sklyanin found the algebraic structure underlying
the above relation, replacing the $L$ operator by
$$
L(\lam) = \sum_{a=0}^3 W_a^L(\lam) \sig^a \tens S^a,
$$
and writing down the relations among $S^a$'s equivalent
to (0.1) [Sk1]. The algebra $Q$ generated by $S^a$'s is
called the Sklyanin algebra. He also constructed series of
representations in [Sk2], among which we use the spin $l$
representation $\rho_l: Q \to$End$(V_l)$, $\dim V_l = 2l+1$.
The algebra $Q$ is a deformation of the universal enveloping
algebra $U(sl(2,\Comp))$, and $\rho_l$ is corresponding
deformation of the spin $l$ representation of $sl(2,\Comp)$.

As the solvability of the $XYZ$ (or 8 vertex) model comes
from the fundamental relation (0.1), it is natural to consider
the generalization of this model by using the above
representations of the Sklyanin algebra, and calculate
its free energy.

\subhead \S1. Description of the model \endsubhead
Now we construct a generalization of the $XYZ$ model by
replacing the $L$ operators of the form (0.2) with
$$
\gathered
L_n(\lam) = \sum_{a=0}^3 W_a^L(\lam) \, \sig^a \tens \rho_l^{(n)} (S^a),\\
\rho_l^{(n)} (S^a) =
1 \tens \cdots \tens 1 \tens
{\underset n-\text{th component} \to{\rho_l(S^a)}}
\tens 1 \tens \cdots 1,
\endgathered
\tag1.1
$$
which act on the space $\Comp^2 \tens V_l^{\tens N}$.
The transition matrix of the higher spin generalization of
the 8 vertex model is defined by:
$$
T(\lam) = L_N(\lam) \cdots L_2(\lam) L_1(\lam).
$$
The transition matrix of the corresponding chain model is
$$
t(\lam) = \tr_{\Comp^2} T(\lam),
$$
which acts on the space $V_l^{\tens N}$.
The free energy of the model under consideration is, by definition,
the Perron-Frobenius eigenvalue of the transfer matrix $t(\lam)$,
$\Lambda_N$, or its thermodynamic limit,
$$
f = \lim_{N \to \infty} \frac1N \ln \Lambda_N.
$$
Similarly to the choice of [TF], we assume that the elliptic modulus
$\tau$ is of the form $\tau = i/t$, where $t>0$ and the parameters $\eta$,
$\lam$ belong to the domain $0<\lam+\eta< \max(\eta, 2l\eta)< {\half}$
(antiferromagnetic region).

In the previous paper [Take], the author showed that the generalized
Bethe Ansatz of Takhtajan and Faddeev is applicable to this model.
According to this result, if $M = Nl$ parameters,
$\lam_1, \ldots \lam_M$, satisfy the Bethe equation,
$$
\left(
\frac{ \theta_{11}(\lam_j - 2l\eta)}{ \theta_{11}(\lam_j + 2l\eta)}
\right)^N
=
\prod_{k=1, k\neq j}^M
\frac{ \theta_{11}(\lam_k - \lam_j - 2\eta)}
{ \theta_{11}(\lam_k -\lam_j + 2\eta)},
\tag1.2
$$
for $j = 1,\ldots, M$, then the transfer matrix $t(\lam)$ has
an eigenvalue
$$
(2\theta_{11}(\lam - 2l\eta))^N
\prod_{k=1}^M
\frac{\theta_{11}(\lam - \lam_k + 2\eta)}{\theta_{11}(\lam - \lam_k)}
+
(2\theta_{11}(\lam + 2l\eta))^N
\prod_{k=1}^M
\frac{\theta_{11}(\lam + \lam_k + 2\eta)}{\theta_{11}(\lam - \lam_k)}.
\tag1.3
$$

\subhead \S2. Thermodynamic limit \endsubhead
Now we take the thermodynamic limit $N \to \infty$.
{}From the results of the $XXX$ and $XXY$ models [Takh], [B] and [S],
we assume that the following {\it string hypothesis} holds also
for this model: assymptotically as $N\to \infty$, the solutions
of the Bethe equation (1.2) corresponding to the ground state
cluster in complexes, so called $2l$-strings:
$$
\{ \lam^0_k + 2\alpha \eta |\, \alpha = -l + \half, \ldots, l + \half
\}_{k = 1,\ldots, N/2}
$$
(assuming that $N$ is even), and centres of these strings $\lam_k^0$
distribute in a purely imaginary interval
$\{ix |\, -1/2t < x < 1/2t\}$ and have a smooth density $\rho(\lam')d\lam'$
as $N\to \infty$ ($\lam' = \lam/\tau$).

Under these hypotheses, the Bethe Ansatz equation (1.2) reduces to
an integral equation
$$
\multline
\sum_{\alpha=0}^{2l-1} \Psi(\lam, (2\alpha+1)\eta) =\\
=
-2\pi\rho(\lam') +
\int^{1/2}_{-1/2} \big(
\sum_{\beta=1}^{2l-1} \Psi(\lam - \mu, 2\beta\eta)
+
\sum_{\beta=0}^{2l-1} \Psi(\lam - \mu, 2(\beta + 1)\eta)
\big)
\rho(\mu')\, d\mu',
\endmultline
\tag2.1
$$
where $\lam' = \lam/\tau$, $\mu' = \mu/\tau$ and
$$
\Psi(\lam, \eta) = \frac\tau i \frac d{d\lam}
\ln \frac{\theta_{11}((\lam-\eta)/\tau;-1/\tau)}
{\theta_{11}((\lam+\eta)/\tau;-1/\tau)}.
$$

We can solve this equation, expanding $\rho(\lam')$ into
Fourier series, and obtain
$$
\rho(\lam') = \sum_{n\in\Int}
\frac{e^{2\pi i n \lam'}}{2\cos 2\pi n \eta'}.
\tag2.2
$$
Using this density of the solutions of the Bethe equation,
we can take the limit of the eigenvalue (1.3) and obtain
the free energy of the model $f= \lim_{N\to\infty}\ln\Lambda_N/N$
as
$$
\multline
e^f \propto \theta_{11}(\lam+2l\eta) e^{2\pi t(\lam-\eta)(2l\eta-\half)}\times
\\
\times
\exp \sum_{n=1}^\infty
\frac
{\sinh 2\pi nt(2l\eta-\half)\,\sinh 2\pi nt(\lam-\eta)}
{n \sinh \pi nt\, \cosh 2\pi nt\eta}=
\\
=
\theta_{11}(\lam+2l\eta)\,
q^{-(\lam-\eta)(2l\eta - \half)}
\frac{\sig(\lam;\eta,l |q)}{\sig(-\lam;\eta,l|q)},
\endmultline
\tag2.3
$$
where $q=\exp(-2\pi t)$ and
$$
\gather
\sig(\lam;\eta,l|q)
=
\prod_{k=0}^\infty
\frac
{\Gamma_q(\lam+4\eta k+2(l+1)\eta)\,\Gamma_q(\lam+4\eta k-2 l   \eta+1)}
{\Gamma_q(\lam+4\eta k+2 l   \eta)\,\Gamma_q(\lam+4\eta k-2(l-1)\eta+1)},
\\
\Gamma_q(x) = \frac{(q;q)_\infty}{(q^x;q)_\infty}(1-q)^{1-x},\qquad
(a;q)_\infty = \prod_{n=0}^\infty (1-aq^n).
\endgather
$$

This result is consistent with that of [TF] which corresponds to
the case $l=\half$.

\medpagebreak
Details and the further results concerning the thermodynamic
properties and the excited states of this model will be
published in the forthcoming paper.

The author expresses his great thanks to Professors~E.~K.~Sklyanin,
H.~J.~de~Vega, M.~Wadati, E.~Date, Kimio~Ueno, M.~Noumi and Dr.~T.~Deguchi
for their comments, advices, interests and encouragement.


\Refs
\widestnumber\key{Takm}

\ref
\key B
\by Babujian H.M.
\jour Phys.lett.A
\vol 90 \yr 1982
\pages 479--482
\endref

\ref
\key M
\by Mumford, D.
\book Tata Lectures on Theta I
\publ Birkh\"auser
\yr1982
\endref

\ref
\key S
\by Sogo K.
\jour Phys.lett.A
\vol 104 \yr 1984
\pages 51--54
\endref

\ref
\key Sk1
\by Sklyanin E.K.
\jour Func. Anal. Appl.
\vol 16-4 \yr 1982
\pages 263--270
\endref

\ref
\key Sk2
\bysame
\jour Func. Anal. Appl.
\vol 17-4 \yr 1983
\pages 273--284
\endref

\ref
\key Take
\by Takebe T.
\jour J.Phys.A
\vol 25 \yr 1992
\pages 1071--1083
\endref

\ref
\key Takh
\by Takhtajan L.A.
\jour Phys.lett.A
\vol 87 \yr 1982
\pages 479--482
\endref

\ref
\key TF
\by Takhtajan L.A., Faddeev L.D.
\jour Uspekhi Mat. Nauk
\vol 34:5 \yr 1979
\pages 13--63
\transl\nofrills English transl. in \jour Russ. Math. Surv.
\vol 34 \yr 1979
\pages 11--68
\endref
\bye